\documentclass[aps,prl,twocolumn,superscriptaddress,groupedaddress]{revtex4} 

\usepackage{graphicx}  % needed for figures
\usepackage{dcolumn}   % needed for some tables
\usepackage{bm}        % for math
\usepackage{amssymb}   % for math
\usepackage{amsmath}
\usepackage{gensymb}
\usepackage[T1]{fontenc}
\usepackage{enumitem}
\usepackage{tikz}
\usepackage[caption=false]{subfig}
\usepackage{graphicx}

\begin{document}

\title{The motion of long levitating drops in tubes in an anti-Bretherton configuration}
\author{Peter Favreau }
\affiliation{
Univ. Lille, CNRS, Centrale Lille, ISEN, Univ. Valenciennes, IEMN UMR 8520, F-59000 Lille, France
}%
\author{Alexis Duchesne}%
\affiliation{
Univ. Lille, CNRS, Centrale Lille, ISEN, Univ. Valenciennes, IEMN UMR 8520, F-59000 Lille, France
}%
 \author{Farzam Zoueshtiagh}%
 \affiliation{
Univ. Lille, CNRS, Centrale Lille, ISEN, Univ. Valenciennes, IEMN UMR 8520, F-59000 Lille, France
}%
\author{Micha\"el Baudoin}%
 \email{michael.baudoin@univ-lille.fr}
\affiliation{
Univ. Lille, CNRS, Centrale Lille, ISEN, Univ. Valenciennes, IEMN UMR 8520, F-59000 Lille, France
}%
\affiliation{Institut Universitaire de France, 1 rue Descartes, 75231 Paris Cedex 05}

\date{\today}% It is always \today, today,

\begin{abstract}
In his seminal paper, Bretherton [J. Fluid Mech., \textbf{10}:166 (1961)] studied the motion of long bubbles in capillary tubes, a situation encountered in many two-phase flow systems. Here, we unveil experimentally and numerically the negative configuration, wherein a long liquid drop formed by the rupture of a liquid plug is stably transported in a capillary tube and surrounded by a flow-induced air cushion.  After a careful theoretical and numerical analysis of the drop formation process, we show that the shape of the drop and lubricating air film is reminiscent of Bretherthon's calculation and can be inferred from an adapted analytical theory. This work opens tremendous perspectives for drop fast transport in microfluidic systems without walls contamination and friction.
\end{abstract}

\maketitle

In his 1961 seminal paper \cite{bretherton1961motion}, Bretherton studied experimentally and theoretically the motion of long bubbles in capillary tubes in the limit of creeping flows, i.e. at low Reynolds and capillary numbers. This apparently academic configuration was first explored in millimetric tubes at very low flow rates or with highly viscous fluids. Then, the emergence of microfluidics \cite{arfm_stone_2004,rmp_squires_2005} led to some renewed interest in Bretherton's theory, owing to its  relevance to various two-phase flow configurations at small scales, including bubbles \cite{bretherton1961motion,park1984two,science_prakash_2007}, plugs \cite{jap_halpern_1998,jfm_howell_2000,baudoin2013airway,magniez2016dynamics,prf_xu_2017,jfm_signe_2018} and foam \cite{el_cantat_2004,csa_denkov_2005,jpcm_hohler_2005} dynamics  in capillary tubes. Moreover, his theory was extended later on to a larger range of flow parameters, e.g. larger capillary and Reynolds numbers \cite{aussillous2000quick,klaseboer2014extended}, more complex tube geometry \cite{wong1995motion,wong1995motion2,hazel2002steady}, or non-Newtonian embedding liquids \cite{jalaal2016long,jfm_laborie_2017}. 

In this paper, we investigate experimentally, theoretically and numerically the negative of Bretherton's configuration, i.e. a long liquid drop moving in an air-filled tube, whose contact with the walls is prevented by a self-induced air cushion. First, we show experimentally and numerically that these levitating drops can be formed by pushing a liquid plug inside a microfluidic channel at a capillary number large enough to induce an inversion of the meniscus with a radius of curvature smaller than the radius of the tube. Second, we develop an analytical model that is able to recover the shape and thickness of the air lubricating film at walls in the limit of low (air) capillary numbers. Finally we explore a larger set of parameters and draw a phase diagram delimiting the regimes in which long levitating droplets are formed. While systems designed to synthesize bubbles or drops in liquids are ubiquitous in microfluidics \cite{apl_anna_2003,m_garstecki_2006,am_nisiako_2006}, the formation of long levitating drop in air has only been reported in superhydrophobic channels \cite{tirandazi2017liquid}, wherein contact with the walls is prevented by the specific surface treatment. This work provides a simple way to generate long drops of controlled length stably propagating in regular tubes, that may serve in digital microfluidics to transport liquids without any wall contamination.

\begin{figure}[ht!]
\begin{center}
\includegraphics[width=8cm]{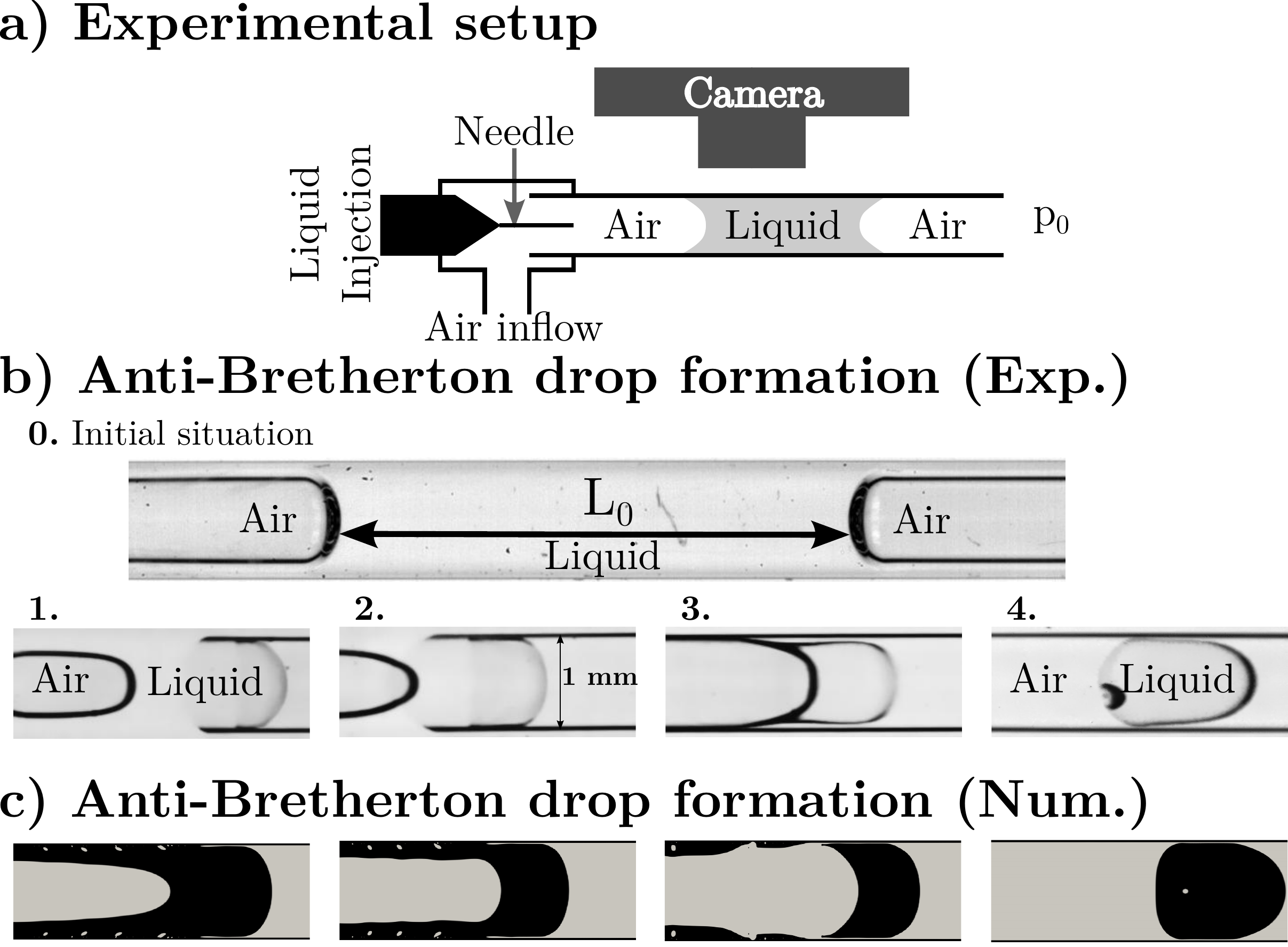}
\caption{Transition from a liquid plug to a long levitation drop. a) Scheme of the experimental device. (b) Snapshot of an experiment: a $100  \,  \mathrm{cST}$ silicone oil liquid plug of length $19  \,  \mathrm{mm}$ is pushed inside a capillary tube of radius $R=0.5  \,  \mathrm{mm}$ by a constant flow of air ($Q = 11  \,  \mathrm{mL/min}$) with a syringe pump, leading to a velocity of the rear interface $U$ of $0.6  \,  \mathrm{m s^{-1}}$ (corresponding to a  capillary number $\mathcal{C}_a^l = \frac{\mu_l U}{\sigma} = 2.8$, with $\mu_l$ the liquid dynamic viscosity and $\sigma$ the surface tension). Image 0 shows the initial plug shape before actuation. Image 1 ($t = 0\, \mathrm{ms}$) shows the deformed rear interface and the inverted front interface when the plug is pushed by a flow of air. Image 2 ($t = 0.6 \, \mathrm{ms}$) shows the evolution of the liquid plug and the appearance of a thin film of air surrounding the liquid at the front of the plug. Image 3 shows ($t = 2.3 \, \mathrm{ms}$) the drop detachment and Image 4 shows the detached long liquid drop stably propagating in the tube.(c) Numerical simulation of the previous experiment with the same flow parameters: black is silicone oil, white is air. }
\label{experimental_device}
\end{center}
\end{figure}

\textbf{Methods and results:} Experimentally, a long levitating drop is synthesized inside a  capillary tube of radius $R = 0.5  \, \mathrm{mm}$ (i) by injecting with a needle a controlled amount of silicone oil in the tube leading to the formation of a liquid plug as depicted on Fig. \ref{experimental_device}.a and (ii) by pushing this liquid plug with a large constant air flow rate $Q$ in the range $\left[1  \,  \mathrm{mL/min}, 50  \,  \mathrm{mL/min} \right]$. The tubes were carefully cleaned prior to experiments with acetone, isopropyl alcohol and dichloromethane. The plug and then droplet evolution is recorded at 10000 or 15000  frames per second with a high speed camera (Photron SA3) mounted on a Leica Z16 macroscope (Fig. \ref{experimental_device}.a and supplementary movie M1). At rest (Image 0), the plug front and rear menisci are two opposite half-sphere tangent to the walls, which fulfill both Young-Laplace law (leading to spherical liquid/air interfaces) and the perfectly wetting condition of silicone oil on the walls (leading to a $0^\circ$ contact angle). Then, a flow rate sufficient to induce an inversion of the curvature of the front interface with a radius of curvature smaller than the tube radius is imposed (Image 1). This leads to the formation of a thin air film surrounding the front part of the liquid plug as evidenced by the greyer color of the liquid (indicating that the liquid is no more contacting the walls). The thin film of air extends progressively downward the plug (Image 2) until it contacts the rear interface of the plug (Image 3) leading to the detachment of the drop, which then propagates stably in the tube (Image 4). The shape of the drop evokes the shape of Bretherton's bubble \cite{bretherton1961motion}, with a phase inversion. Note that to improve the repeatability of the experiments and ease the comparison with the simulations (avoid contact line problems), the walls were prewetted  prior to the experiments, by injecting a liquid plug inside the channel and pushing it with a low constant flow rate $Q = 4.5 \, \mathrm{\mu L/min} $ (corresponding to a low capillary number $\mathcal{C}_a^l \approx 4.6\times 10^{-4}$ leading to the deposition on the walls of a thin liquid film of controlled thickness \cite{bretherton1961motion,magniez2016dynamics}  $h_p =0.643 R \left(3 \mathcal{C}_a^l \right)^{2/3} \approx 4 \, \mathrm{\mu m}$, with $R$ the radius of the capillary tube, $\mathcal{C}_a^l = \frac{\mu_l U}{\sigma}$ the capillary number of the liquid comparing viscous effects to surface tension ones, $\mu_l$ the liquid dynamic viscosity and $U$ the speed of the rear interface of the plug.

Numerically, the same dynamics (Fig. \ref{experimental_device}.c) is simulated in a 2D axisymmetric configuration with a Volume of Fluid Method (VOF) \cite{hirt1981volume} implemented in the open-source code \emph{OpenFOAM} solving the following set of equations:
\begin{eqnarray}
& \mathbf{\nabla}\cdot\mathbf{u}=0 \nonumber \\
& \partial_t\rho\mathbf{u}+\mathbf{\nabla}\cdot\left(\rho\mathbf{u}\otimes\mathbf{u}\right)=\rho\mathbf{g}-\mathbf{\nabla}\left(p\right)+\mu\mathbf{\Delta}\left(\mathbf{u}\right)+\sigma\kappa\mathbf{n}\delta_{S} \nonumber  \\
& \partial_t\alpha+\mathbf{\nabla}.\left(\alpha\mathbf{u}\right)=0 \nonumber \\ 
& \rho=\alpha\rho_{l}+\left(1-\alpha\right)\rho_{g}\,\,\,\,;\,\,\,\,\mu=\alpha\mu_{l}+\left(1-\alpha\right)\mu_{g} \nonumber
\end{eqnarray}
where  $\sigma$ is the surface tension, $\mu_i$ and $\rho_i$ the viscosity and density of the phase $i$, ($i = l$ for the liquid and $g$ for the gas), $\mathbf{u}$ the fluid velocity, $p$ the dynamic pressure and $\mathbf{g}$ the gravitational  acceleration. $ \alpha $ is a phase marker equal to $0$ in the liquid and 1 in the gas. The interface between the two fluids thus corresponds to $ \alpha \in \left] 0 \; ; 1 \right[$. The term taking into account the effects of surface tension is $ \sigma \kappa \mathbf{n} \delta_{S}$, with $\delta$ the chronecker symbol, equal to 1 on the interface. This modeling was introduced by Brackbill et al. \cite{brackbill1992continuum} and called the Continuum Surface Force (CSF) model. Since the Volume of Fluid method is known to generate parasitic currents \cite{abadie2015combined,francois2006balanced,harvie2006analysis}, i.e. spurious hydrodynamic vortices close to the interface, a restrictive time step was imposed to guarantee the stability of this parasitic flow \cite{deshpande2012evaluating} \begingroup\small$\Delta t \leq \max\left( 0.1\sqrt{\frac{\rho \Delta x^3}{\sigma}}\; ; \; 10 \frac{\mu \Delta x}{\sigma} \right)$\endgroup, with $\Delta x$ the characteristic length of a mesh cell. We also implemented an adaptive mesh in the code. Indeed, the correct calculation of the flow close to the walls and interfaces requires a refined mesh for the sake of precision. Without adaptive mesh, the calculation cost would be prohibitive.

\begin{figure}[h!]
\begin{centering}
\includegraphics[width=8cm]{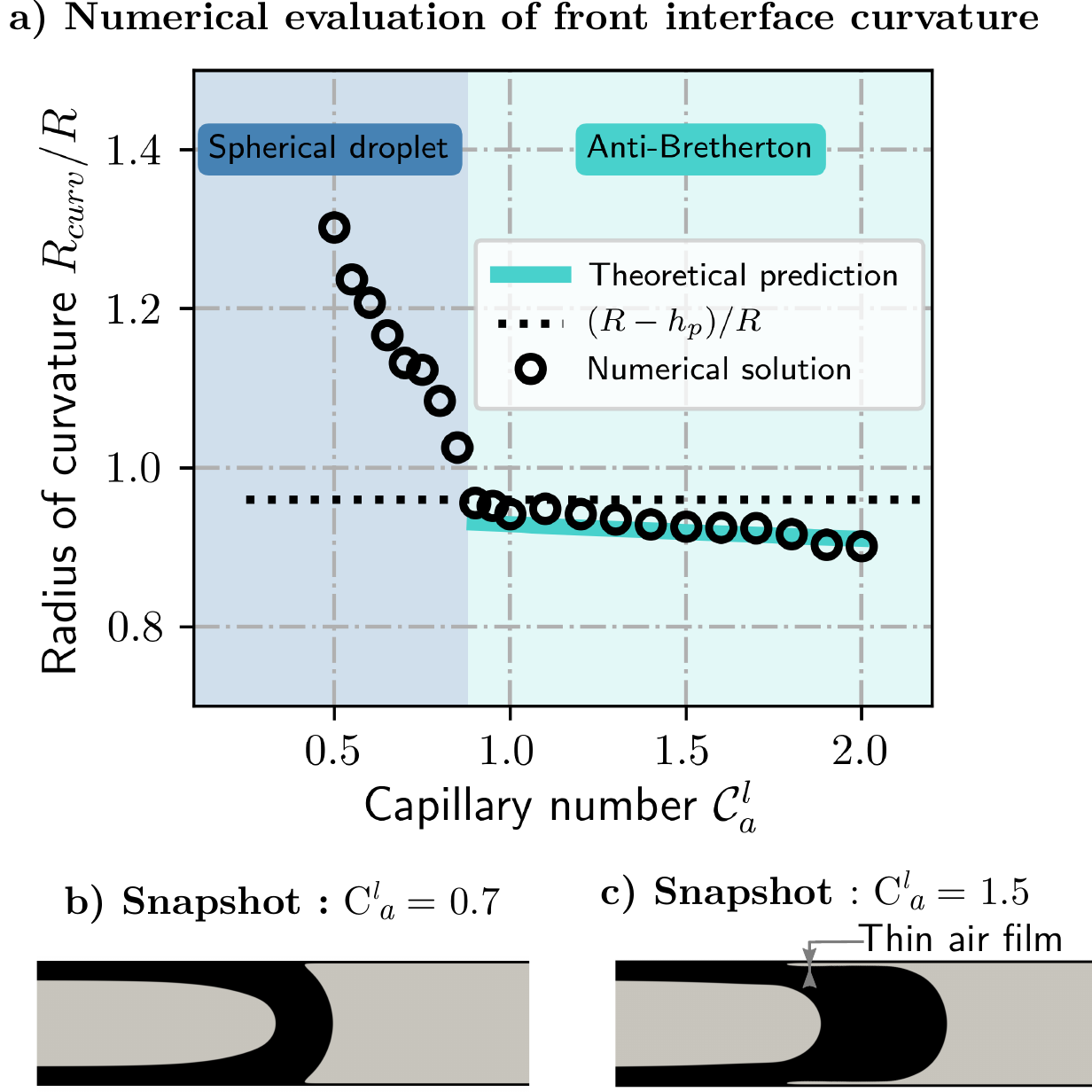}
\end{centering}
\caption{(a) Numerical evaluation of the front interface curvature $R_{curv}$ of the front interface of a liquid plug divided by the tube radius $R$ as a function of the capillary number $\mathcal{C}_a^l$ (at constant Ohnesorge number $\mathcal{O}_{h}=0.4$). When the radius of curvature exceeds the tube radius (snapshot (b)), spherical droplets are formed when the plug ruptures. When it becomes smaller than the tube radius (above a critical capillary number), a thin film of air appears between the plug and the walls (snapshot (c)) leading to the formation of an anti-Bretherton drop. (b) Snapshot of numerical simulation: $\mathcal{C}_a^l=0.7$, $\mathcal{O}_{h}=0.4$
(c) Snapshot of numerical simulation: $\mathcal{C}_a^l=1.5$, $\mathcal{O}_{h}=0.4$}
\label{curvradius}
\end{figure}

\textbf{Transition from liquid plug to long levitating drop:}  The dynamics of liquid plugs in perfectly wetting channels has been widely studied experimentally \cite{baudoin2013airway,magniez2016dynamics,ijmf_signe_2019}, numerically \cite{fujioka2005steady,pof_fujioka_2008,hassan2011adaptive} and theoretically \cite{jap_halpern_1998,jfm_howell_2000,jcis_jensen_2000,baudoin2013airway,prf_xu_2017,jfm_signe_2018}. From a theoretical perspective, a liquid plug can be seen as a bridge of liquid trapped between two semi-infinite air bubbles. Hence, the laws of deformation of the front (respectively rear) interface of a liquid plug can be inferred from corresponding laws derived for the deformation of the rear (respectively front) interface of long bubbles \cite{bretherton1961motion, ratulowski1989transport, park1984two,  ajaev2006modeling}. Nevertheless, most studies have been conducted in the analytically tractable low or intermediate \cite{aussillous2000quick,klaseboer2014extended} capillary number limit, wherein the curvature sign of the meniscus is not changed. 

Hoffman \cite{hoffman1975study} studied experimentally the evolution of the front meniscus of an advancing liquid finger on a large range of liquid capillary number (ranging from $\approx 4 \times 10^{-5}$ to $\approx 35$). He reported an evolution of the apparent contact angle from $\approx 5^\circ$ to $\approx 180^\circ$, corresponding to an inversion of the front meniscus curvature from $C \approx -2/R$ to $C \approx 2/R$ (with $R$ the radius of the tube). This behaviour was rationalized later on by Boender et al. \cite{boender1991approximate} with approximate analytical models. Here we show both experimentally (Fig. \ref{experimental_device}b) and numerically (Fig \ref{curvradius}a,c) that when the capillary number exceeds the critical capillary number $ \mathcal{C}_{a,crit}^l $ at which the radius of cuvature becomes equal to the tube radius $R$ (or more precisely $R - h_p$ in this paper owing to the existence of a prewetting film), this latter continues to decrease (while more slowly) through the appearance of a thin film of air which fills the gap between the front meniscus and the walls (Fig. \ref{curvradius}c) . The appearance of this film of air is the key ingredient toward the formation of long levitating drops from the fast dynamics of liquid plugs. Indeed, this film of air extends progressively backward (Fig. \ref{experimental_device}b) until it reaches the back of the drop, hence provoking its detachment. Interestingly, this deposition of a thin film of air behind the drop front interface is reminiscent of the deposition of a thin film of liquid behind the front interface of a bubble predicted by Bretherthon's theory, but with a phase inversion.
\begin{figure}[h!]
\begin{center}
\includegraphics[width=8cm]{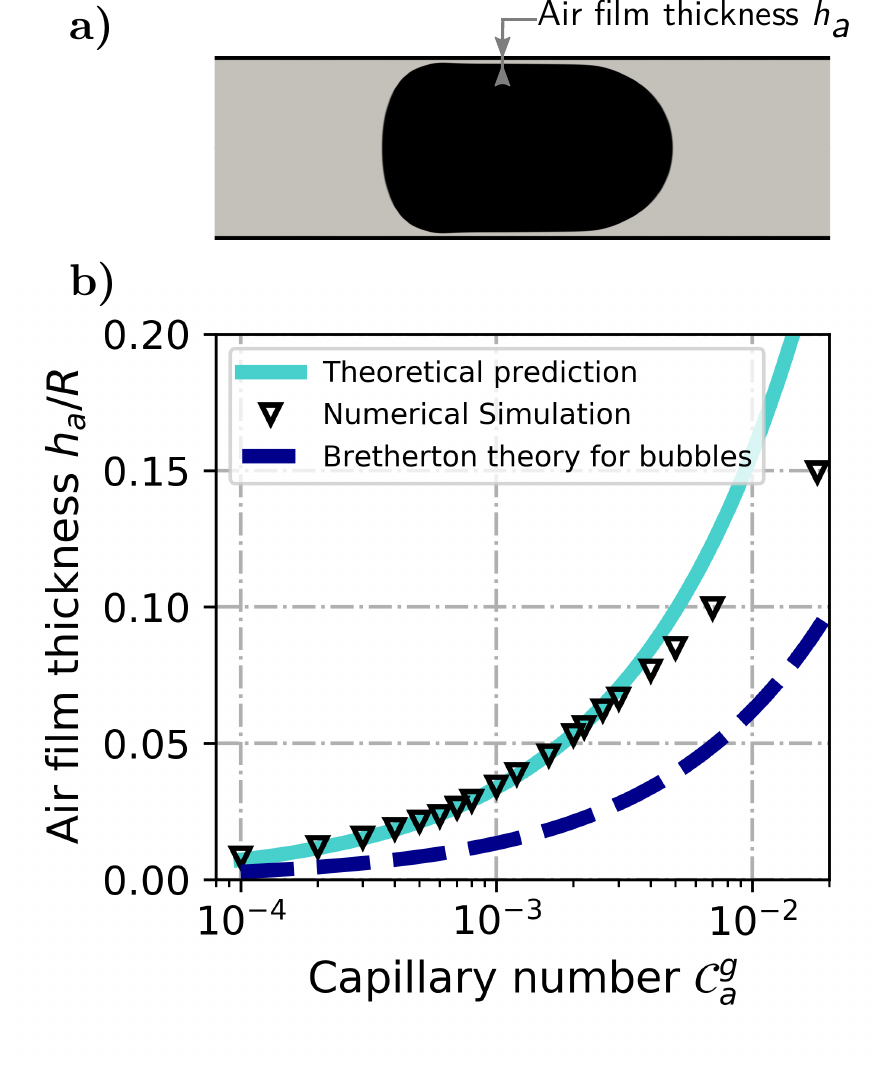}
\caption{a. Simulation showing a long levitating drop moving in a capillary tube and separated from the walls by a thin air film of thickness $h_a$ b. Evaluation of the thickness of the air film $h_a$ as a function of the air capillary number $\mathcal{C}_a^g$. Triangles: numerical simulation. Cyan continuous line: adapted Bretherton theory for levitating drops. Blue dashed line: Bretherton's theory for bubbles.}
\label{courbe_loi_depot}
\end{center}
\end{figure}

\textbf{Thin air film thickness prediction: anti-Bretherton.} Since these two configurations share some similarities, we derived a model to evaluate the thickness of the air film as a function of the capillary number by following a similar procedure as the one proposed by Bretherton. First, it is important to stress out that two capillary numbers are involved in this problem: one based on the gaseous phase $ \mathcal{C}_a^g=  \mu_g U / \sigma $ and one based on the liquid $ \mathcal{C}_a^l = \mu_l U / \sigma $ of course proportional to each other, $\mathcal{C}_a^g= \mu_g / \mu_l \, \mathcal{C}_a^l$, with a constant depending on the gas and liquid viscosity ratio. The condition for the formation of an Anti-Bretherton drop from a liquid plug imposes that $ \mathcal{C}_a^l > \mathcal{C}_{a,crit}^l \sim \mathcal{O} (1)$, since viscous stresses must overcome surface tension to induce an inversion of the front interface up to a point wherein the radius of curvature becomes smaller than the radius of the tube. On the other hand, the thickness of the film of air appearing behind the drop front interface relies on the capillary number in the air $\mathcal{C}_a^g$. This capillary number remains small ($\mathcal{C}_a^g\in \left[ 10^{-4} \; ; \; 10^{-2}\right]$) in all the simulations and experiments provided in this paper, as well as the Reynolds number associated to the flow in the air film ($\mathcal{R}_e^g \in \left[10^{-3} \; ; \; 2 \right]$). Hence, the classic lubrication approximation can be used to describe the flow in the thin air film:
\begin{equation}
\begin{cases}
\begin{array}{c}
\partial_{x}p_g=\mu_g\partial_{yy}^{2}u_{x}\\
\partial_{y}p_g=0
\end{array}\end{cases}\Rightarrow \;\;\frac{dp}{dx}=\mu\frac{d^{2}u_{x}}{dy^{2}}
\end{equation}
with $u_{x}$ and $p_g$ the longitudinal velocity and pressure in the air film. Owing to the large difference of viscosity between the fluid and the air, the thin prewetting film of liquid on the walls can be considered at rest and the velocity inside the detaching drop can be considered as constant. This approximation is confirmed by the numerical simulations. Hence, in the drop frame of reference, the boundary conditions at the walls (or more precisely a the prewetting film surface) becomes: $u_{x} \left(y = 0 \right) = - U$ with $ U $ the drop speed, while at the interface between the film and the drop, i.e. at $y = h(x) $, we have $u_{x}(h(x))=0$ where $h(x)$ denotes the thickness of the air film. The normal stress balance at the drop interface gives: $ p = - \sigma \left (\frac{1}{R} + \frac{d ^ {2} h(x)}{dx ^{2} } \right) $, with $ p = p_g - p_l $, where $ p_g $ and $ p_l $ are the pressures in the gas and liquid phase respectively. Finally, the mass conservation in the air film gives: $\int_{0}^{h(x)} u_{x} dy=-U H$, with $H$, the constant thickness of the film far from the drop front meniscus. The only difference with the equations derived by Bretherton is the boundary condition at $y = h(x) $, which for a drop (present case) is an adherence condition $u_{x}(h(x))=0$, while for a bubble is a zero stress condition. The combination of these equations with the change of variable $x=H\left(12 \, \mathcal{C}_a^g \right)^{-1/3}\xi$ and $h(x)=H \psi$, leads to the celebrated Landau-Levich equation \cite{bretherton1961motion,landau2012dragging}:
\begin{equation}
\frac{d^{3}\psi}{d\xi^{3}}=\frac{\psi-1}{\psi^{3}}
\end{equation}
This equation coincides with the one obtained by Bretherton, except from the coefficient in the change of variable $x=H\left(12 \, \mathcal{C}_a^g\right)^{-1/3}\xi$, which is $12$ instead of $3$ for a bubble. Hence, we can infer the thickness of the film of air from the solution obtained by Bretherton by a simple change of cofficient:
\begin{equation}
\frac{h_a}{R}=0.643\left(12 \; \mathcal{C}_a^g\right)^{2/3}
\label{loi_depot_anti_bretherton}
\end{equation}
The validity of this analytical expression was verified through the comparison with numerical simulations of the dynamics of long drops in capillary tubes resulting from the rupture of a liquid plug at high liquid capillary numbers (Fig. \ref{courbe_loi_depot}). The thickness of the film was measured in the flat part of the drop away from the menisci, once it reaches a stable shape. The results of the simulation are represented on Fig. \ref{courbe_loi_depot} (triangles) and compared with both the present analytical expression (cyan continuous line)  and the one obtained by Bretherton for bubbles (purple dashed line). The analytical expression obtained is in excellent agreement with the simulations in the limit of low capillary numbers $\mathcal{C}_a^g < 3 \times 10^{-3}$ and differs, as expected, from the simulations for larger capillary numbers \cite{aussillous2000quick}.

\begin{figure}[h!]
\begin{centering}
\includegraphics[width=8cm]{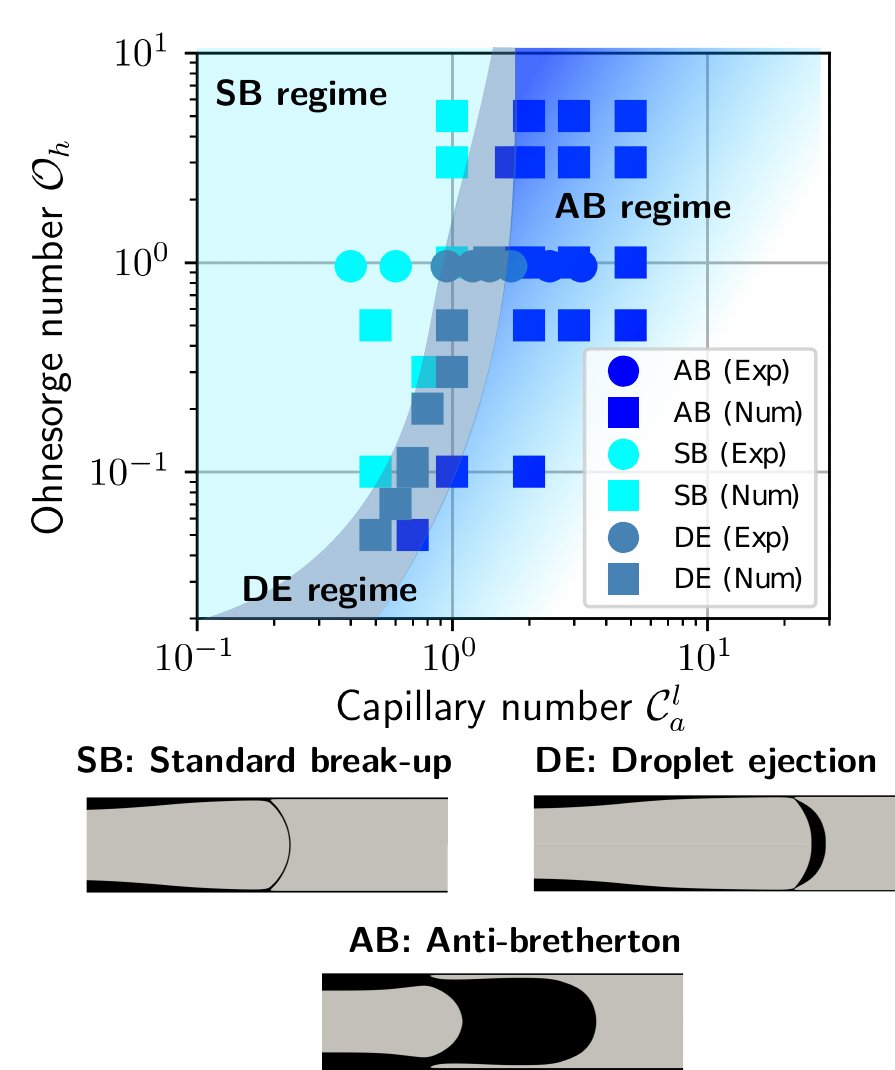}
\end{centering}
\caption{Phase diagram summarizing the regimes observed experimentally and numerically for different Capillary and Ohnesorge numbers. Standard Break-up (SB): the plug breaks from the center with no plug formation. Droplet ejection (DE): a spherical drop is ejected at the center of the channel. Anti-Bretherton (AB) a long levitating drop separated from the walls by an air cushion is formed (see Movie M2).}
\label{figure4}
\end{figure}

\textbf{Phase diagram:} We further investigated experimentally and numerically the regimes leading to the formation of long anti-Bretherton drops, on a large set of parameters. Since the 3 effects at stake are viscous, inertial and capillary effects, a phase diagram can be plotted as a function of two dimensionless numbers: the capillary number $\mathcal{C}_a^l = \mu_l U / \sigma$ and the Ohnesorge number $\mathcal{O}_h = \mu_l / \sqrt{\rho_l \sigma R}$ (see Fig. \ref{figure4}). The Ohnesorge number compares viscous effects stabilizing an interface to the geometric average of desabilizing inertial and capillary effects. In the range of capillary number sufficient to have an inversion of the front interface curvature, 3 regimes can be observed (see Fig. \ref{figure4} and movie M2): (i) A regime named "Standard Break-up" (SB) wherein the liquid plug becomes thinner and thinner until it breaks with no droplet production. This regime occurs when the radius of curvature of the front interface remains larger than the tube radius. (ii) An intermediate regime named "Droplet Ejection" (DE) wherein a small drop is ejected occuring when the front interface radius of curvature is almost equal to the tube radius. And finally (iii) the "Anti-Bretherton" large drop production regime leading to the production of long droplet separated from the tube walls by a thin air film. This regime (as discussed before) occurs when the radius of curvature of the front interface becomes smaller than the tube radius.

\textbf{Conclusion:} In this paper we study experimentally, theoretically and numerically the dynamics of large drops levitating in a capillary tube, produced by the rupture of a liquid plug. We show that the production of these large drops occurs when the radius of curvature of the front interface of the liquid plug becomes smaller than the tube radius leading to the appearance of a thin air film between the plug and the walls, propagating downward the plug. The drops shape are reminiscent from the one of long bubble propagating in a liquid filled tube and can be inferred from Bretherton's model with inverted phases and adapted boundary conditions. This work opens perspective for high speed transport of droplets in tubes with no wall contamination.

\bibliographystyle{apsrev4-2}
%\bibliography{Bibliographie/biblio_complete.bib}

\begin{thebibliography}{41}%
\makeatletter
\providecommand \@ifxundefined [1]{%
 \@ifx{#1\undefined}
}%
\providecommand \@ifnum [1]{%
 \ifnum #1\expandafter \@firstoftwo
 \else \expandafter \@secondoftwo
 \fi
}%
\providecommand \@ifx [1]{%
 \ifx #1\expandafter \@firstoftwo
 \else \expandafter \@secondoftwo
 \fi
}%
\providecommand \natexlab [1]{#1}%
\providecommand \enquote  [1]{``#1''}%
\providecommand \bibnamefont  [1]{#1}%
\providecommand \bibfnamefont [1]{#1}%
\providecommand \citenamefont [1]{#1}%
\providecommand \href@noop [0]{\@secondoftwo}%
\providecommand \href [0]{\begingroup \@sanitize@url \@href}%
\providecommand \@href[1]{\@@startlink{#1}\@@href}%
\providecommand \@@href[1]{\endgroup#1\@@endlink}%
\providecommand \@sanitize@url [0]{\catcode `\\12\catcode `\$12\catcode
  `\&12\catcode `\#12\catcode `\^12\catcode `\_12\catcode `\%12\relax}%
\providecommand \@@startlink[1]{}%
\providecommand \@@endlink[0]{}%
\providecommand \url  [0]{\begingroup\@sanitize@url \@url }%
\providecommand \@url [1]{\endgroup\@href {#1}{\urlprefix }}%
\providecommand \urlprefix  [0]{URL }%
\providecommand \Eprint [0]{\href }%
\providecommand \doibase [0]{https://doi.org/}%
\providecommand \selectlanguage [0]{\@gobble}%
\providecommand \bibinfo  [0]{\@secondoftwo}%
\providecommand \bibfield  [0]{\@secondoftwo}%
\providecommand \translation [1]{[#1]}%
\providecommand \BibitemOpen [0]{}%
\providecommand \bibitemStop [0]{}%
\providecommand \bibitemNoStop [0]{.\EOS\space}%
\providecommand \EOS [0]{\spacefactor3000\relax}%
\providecommand \BibitemShut  [1]{\csname bibitem#1\endcsname}%
\let\auto@bib@innerbib\@empty
%</preamble>
\bibitem [{\citenamefont {Bretherton}(1961)}]{bretherton1961motion}%
  \BibitemOpen
  \bibfield  {author} {\bibinfo {author} {\bibfnamefont {F.}~\bibnamefont
  {Bretherton}},\ }\href@noop {} {\bibfield  {journal} {\bibinfo  {journal}
  {Journal of Fluid Mechanics}\ }\textbf {\bibinfo {volume} {10}},\ \bibinfo
  {pages} {166} (\bibinfo {year} {1961})}\BibitemShut {NoStop}%
\bibitem [{\citenamefont {Stone}\ \emph {et~al.}(2004)\citenamefont {Stone},
  \citenamefont {Stroock},\ and\ \citenamefont {Ajdari}}]{arfm_stone_2004}%
  \BibitemOpen
  \bibfield  {author} {\bibinfo {author} {\bibfnamefont {H.}~\bibnamefont
  {Stone}}, \bibinfo {author} {\bibfnamefont {A.}~\bibnamefont {Stroock}},\
  and\ \bibinfo {author} {\bibfnamefont {A.}~\bibnamefont {Ajdari}},\
  }\href@noop {} {\bibfield  {journal} {\bibinfo  {journal} {Annu. Rev. Fluid
  Mech.}\ }\textbf {\bibinfo {volume} {36}},\ \bibinfo {pages} {381} (\bibinfo
  {year} {2004})}\BibitemShut {NoStop}%
\bibitem [{\citenamefont {Squires}\ and\ \citenamefont
  {Quake}(2005)}]{rmp_squires_2005}%
  \BibitemOpen
  \bibfield  {author} {\bibinfo {author} {\bibfnamefont {T.}~\bibnamefont
  {Squires}}\ and\ \bibinfo {author} {\bibfnamefont {S.}~\bibnamefont
  {Quake}},\ }\href@noop {} {\bibfield  {journal} {\bibinfo  {journal} {Rev.
  Mod. Phys.}\ }\textbf {\bibinfo {volume} {77}},\ \bibinfo {pages} {977}
  (\bibinfo {year} {2005})}\BibitemShut {NoStop}%
\bibitem [{\citenamefont {Park}\ and\ \citenamefont
  {Homsy}(1984)}]{park1984two}%
  \BibitemOpen
  \bibfield  {author} {\bibinfo {author} {\bibfnamefont {C.-W.}\ \bibnamefont
  {Park}}\ and\ \bibinfo {author} {\bibfnamefont {G.}~\bibnamefont {Homsy}},\
  }\href@noop {} {\bibfield  {journal} {\bibinfo  {journal} {Journal of Fluid
  Mechanics}\ }\textbf {\bibinfo {volume} {139}},\ \bibinfo {pages} {291}
  (\bibinfo {year} {1984})}\BibitemShut {NoStop}%
\bibitem [{\citenamefont {Prakash}\ and\ \citenamefont
  {Gershenfeld}(2007)}]{science_prakash_2007}%
  \BibitemOpen
  \bibfield  {author} {\bibinfo {author} {\bibfnamefont {M.}~\bibnamefont
  {Prakash}}\ and\ \bibinfo {author} {\bibfnamefont {N.}~\bibnamefont
  {Gershenfeld}},\ }\href@noop {} {\bibfield  {journal} {\bibinfo  {journal}
  {Science}\ }\textbf {\bibinfo {volume} {315}},\ \bibinfo {pages} {832}
  (\bibinfo {year} {2007})}\BibitemShut {NoStop}%
\bibitem [{\citenamefont {Halpern}\ \emph {et~al.}(1998)\citenamefont
  {Halpern}, \citenamefont {Jensen},\ and\ \citenamefont
  {Grotberg}}]{jap_halpern_1998}%
  \BibitemOpen
  \bibfield  {author} {\bibinfo {author} {\bibfnamefont {D.}~\bibnamefont
  {Halpern}}, \bibinfo {author} {\bibfnamefont {O.}~\bibnamefont {Jensen}},\
  and\ \bibinfo {author} {\bibfnamefont {J.}~\bibnamefont {Grotberg}},\
  }\href@noop {} {\bibfield  {journal} {\bibinfo  {journal} {J. Appl. Phys.}\
  }\textbf {\bibinfo {volume} {85}},\ \bibinfo {pages} {333} (\bibinfo {year}
  {1998})}\BibitemShut {NoStop}%
\bibitem [{\citenamefont {Howell}\ \emph {et~al.}(2000)\citenamefont {Howell},
  \citenamefont {Waters},\ and\ \citenamefont {Grotberg}}]{jfm_howell_2000}%
  \BibitemOpen
  \bibfield  {author} {\bibinfo {author} {\bibfnamefont {P.}~\bibnamefont
  {Howell}}, \bibinfo {author} {\bibfnamefont {S.}~\bibnamefont {Waters}},\
  and\ \bibinfo {author} {\bibfnamefont {J.}~\bibnamefont {Grotberg}},\
  }\href@noop {} {\bibfield  {journal} {\bibinfo  {journal} {J. Fluid Mech.}\
  }\textbf {\bibinfo {volume} {406}},\ \bibinfo {pages} {309} (\bibinfo {year}
  {2000})}\BibitemShut {NoStop}%
\bibitem [{\citenamefont {Baudoin}\ \emph {et~al.}(2013)\citenamefont
  {Baudoin}, \citenamefont {Song}, \citenamefont {Manneville},\ and\
  \citenamefont {Baroud}}]{baudoin2013airway}%
  \BibitemOpen
  \bibfield  {author} {\bibinfo {author} {\bibfnamefont {M.}~\bibnamefont
  {Baudoin}}, \bibinfo {author} {\bibfnamefont {Y.}~\bibnamefont {Song}},
  \bibinfo {author} {\bibfnamefont {P.}~\bibnamefont {Manneville}},\ and\
  \bibinfo {author} {\bibfnamefont {C.~N.}\ \bibnamefont {Baroud}},\
  }\href@noop {} {\bibfield  {journal} {\bibinfo  {journal} {Proceedings of the
  National Academy of Sciences}\ }\textbf {\bibinfo {volume} {110}},\ \bibinfo
  {pages} {859} (\bibinfo {year} {2013})}\BibitemShut {NoStop}%
\bibitem [{\citenamefont {Magniez}\ \emph {et~al.}(2016)\citenamefont
  {Magniez}, \citenamefont {Baudoin}, \citenamefont {Liu},\ and\ \citenamefont
  {Zoueshtiagh}}]{magniez2016dynamics}%
  \BibitemOpen
  \bibfield  {author} {\bibinfo {author} {\bibfnamefont {J.}~\bibnamefont
  {Magniez}}, \bibinfo {author} {\bibfnamefont {M.}~\bibnamefont {Baudoin}},
  \bibinfo {author} {\bibfnamefont {C.}~\bibnamefont {Liu}},\ and\ \bibinfo
  {author} {\bibfnamefont {F.}~\bibnamefont {Zoueshtiagh}},\ }\href@noop {}
  {\bibfield  {journal} {\bibinfo  {journal} {Soft Matter}\ }\textbf {\bibinfo
  {volume} {12}},\ \bibinfo {pages} {8710} (\bibinfo {year}
  {2016})}\BibitemShut {NoStop}%
\bibitem [{\citenamefont {Xu}\ and\ \citenamefont
  {Jensens}(2017)}]{prf_xu_2017}%
  \BibitemOpen
  \bibfield  {author} {\bibinfo {author} {\bibfnamefont {F.}~\bibnamefont
  {Xu}}\ and\ \bibinfo {author} {\bibfnamefont {O.}~\bibnamefont {Jensens}},\
  }\href@noop {} {\bibfield  {journal} {\bibinfo  {journal} {Phys. Rev.
  Fluids}\ }\textbf {\bibinfo {volume} {2}},\ \bibinfo {pages} {094004}
  (\bibinfo {year} {2017})}\BibitemShut {NoStop}%
\bibitem [{\citenamefont {Sign\'{e}~Mamba}\ \emph {et~al.}(2018)\citenamefont
  {Sign\'{e}~Mamba}, \citenamefont {Magniez}, \citenamefont {Zoueshtiagh},\
  and\ \citenamefont {Baudoin}}]{jfm_signe_2018}%
  \BibitemOpen
  \bibfield  {author} {\bibinfo {author} {\bibfnamefont {S.}~\bibnamefont
  {Sign\'{e}~Mamba}}, \bibinfo {author} {\bibfnamefont {J.}~\bibnamefont
  {Magniez}}, \bibinfo {author} {\bibfnamefont {F.}~\bibnamefont
  {Zoueshtiagh}},\ and\ \bibinfo {author} {\bibfnamefont {M.}~\bibnamefont
  {Baudoin}},\ }\href@noop {} {\bibfield  {journal} {\bibinfo  {journal} {J.
  Fluid Mech.}\ }\textbf {\bibinfo {volume} {838}},\ \bibinfo {pages} {165}
  (\bibinfo {year} {2018})}\BibitemShut {NoStop}%
\bibitem [{\citenamefont {Cantat}\ \emph {et~al.}(2004)\citenamefont {Cantat},
  \citenamefont {Kern},\ and\ \citenamefont {Delannay}}]{el_cantat_2004}%
  \BibitemOpen
  \bibfield  {author} {\bibinfo {author} {\bibfnamefont {I.}~\bibnamefont
  {Cantat}}, \bibinfo {author} {\bibfnamefont {N.}~\bibnamefont {Kern}},\ and\
  \bibinfo {author} {\bibfnamefont {R.}~\bibnamefont {Delannay}},\ }\href@noop
  {} {\bibfield  {journal} {\bibinfo  {journal} {Europhys. Lett.}\ }\textbf
  {\bibinfo {volume} {65}},\ \bibinfo {pages} {726} (\bibinfo {year}
  {2004})}\BibitemShut {NoStop}%
\bibitem [{\citenamefont {Denkov}\ \emph {et~al.}(2005)\citenamefont {Denkov},
  \citenamefont {Subramanian}, \citenamefont {Gurovich},\ and\ \citenamefont
  {Lips}}]{csa_denkov_2005}%
  \BibitemOpen
  \bibfield  {author} {\bibinfo {author} {\bibfnamefont {N.}~\bibnamefont
  {Denkov}}, \bibinfo {author} {\bibfnamefont {V.}~\bibnamefont {Subramanian}},
  \bibinfo {author} {\bibfnamefont {D.}~\bibnamefont {Gurovich}},\ and\
  \bibinfo {author} {\bibfnamefont {A.}~\bibnamefont {Lips}},\ }\href@noop {}
  {\bibfield  {journal} {\bibinfo  {journal} {Coll. Surf. A}\ }\textbf
  {\bibinfo {volume} {263}},\ \bibinfo {pages} {129} (\bibinfo {year}
  {2005})}\BibitemShut {NoStop}%
\bibitem [{\citenamefont {H\"{o}hler}\ and\ \citenamefont
  {Cohen-Addad}(2005)}]{jpcm_hohler_2005}%
  \BibitemOpen
  \bibfield  {author} {\bibinfo {author} {\bibfnamefont {R.}~\bibnamefont
  {H\"{o}hler}}\ and\ \bibinfo {author} {\bibfnamefont {S.}~\bibnamefont
  {Cohen-Addad}},\ }\href@noop {} {\bibfield  {journal} {\bibinfo  {journal}
  {J. Phys.: Condens. Matter}\ }\textbf {\bibinfo {volume} {17}},\ \bibinfo
  {pages} {1041} (\bibinfo {year} {2005})}\BibitemShut {NoStop}%
\bibitem [{\citenamefont {Aussillous}\ and\ \citenamefont
  {Qu{\'e}r{\'e}}(2000)}]{aussillous2000quick}%
  \BibitemOpen
  \bibfield  {author} {\bibinfo {author} {\bibfnamefont {P.}~\bibnamefont
  {Aussillous}}\ and\ \bibinfo {author} {\bibfnamefont {D.}~\bibnamefont
  {Qu{\'e}r{\'e}}},\ }\href@noop {} {\bibfield  {journal} {\bibinfo  {journal}
  {Physics of fluids}\ }\textbf {\bibinfo {volume} {12}},\ \bibinfo {pages}
  {2367} (\bibinfo {year} {2000})}\BibitemShut {NoStop}%
\bibitem [{\citenamefont {Klaseboer}\ \emph {et~al.}(2014)\citenamefont
  {Klaseboer}, \citenamefont {Gupta},\ and\ \citenamefont
  {Manica}}]{klaseboer2014extended}%
  \BibitemOpen
  \bibfield  {author} {\bibinfo {author} {\bibfnamefont {E.}~\bibnamefont
  {Klaseboer}}, \bibinfo {author} {\bibfnamefont {R.}~\bibnamefont {Gupta}},\
  and\ \bibinfo {author} {\bibfnamefont {R.}~\bibnamefont {Manica}},\
  }\href@noop {} {\bibfield  {journal} {\bibinfo  {journal} {Physics of
  Fluids}\ }\textbf {\bibinfo {volume} {26}},\ \bibinfo {pages} {032107}
  (\bibinfo {year} {2014})}\BibitemShut {NoStop}%
\bibitem [{\citenamefont {Wong}\ \emph
  {et~al.}(1995{\natexlab{a}})\citenamefont {Wong}, \citenamefont {Radke},\
  and\ \citenamefont {Morris}}]{wong1995motion}%
  \BibitemOpen
  \bibfield  {author} {\bibinfo {author} {\bibfnamefont {H.}~\bibnamefont
  {Wong}}, \bibinfo {author} {\bibfnamefont {C.}~\bibnamefont {Radke}},\ and\
  \bibinfo {author} {\bibfnamefont {S.}~\bibnamefont {Morris}},\ }\href@noop {}
  {\bibfield  {journal} {\bibinfo  {journal} {Journal of Fluid Mechanics}\
  }\textbf {\bibinfo {volume} {292}},\ \bibinfo {pages} {71} (\bibinfo {year}
  {1995}{\natexlab{a}})}\BibitemShut {NoStop}%
\bibitem [{\citenamefont {Wong}\ \emph
  {et~al.}(1995{\natexlab{b}})\citenamefont {Wong}, \citenamefont {Radke},\
  and\ \citenamefont {Morris}}]{wong1995motion2}%
  \BibitemOpen
  \bibfield  {author} {\bibinfo {author} {\bibfnamefont {H.}~\bibnamefont
  {Wong}}, \bibinfo {author} {\bibfnamefont {C.}~\bibnamefont {Radke}},\ and\
  \bibinfo {author} {\bibfnamefont {S.}~\bibnamefont {Morris}},\ }\href@noop {}
  {\bibfield  {journal} {\bibinfo  {journal} {Journal of Fluid Mechanics}\
  }\textbf {\bibinfo {volume} {292}},\ \bibinfo {pages} {95} (\bibinfo {year}
  {1995}{\natexlab{b}})}\BibitemShut {NoStop}%
\bibitem [{\citenamefont {Hazel}\ and\ \citenamefont
  {Heil}(2002)}]{hazel2002steady}%
  \BibitemOpen
  \bibfield  {author} {\bibinfo {author} {\bibfnamefont {A.~L.}\ \bibnamefont
  {Hazel}}\ and\ \bibinfo {author} {\bibfnamefont {M.}~\bibnamefont {Heil}},\
  }\href@noop {} {\bibfield  {journal} {\bibinfo  {journal} {Journal of Fluid
  Mechanics}\ }\textbf {\bibinfo {volume} {470}},\ \bibinfo {pages} {91}
  (\bibinfo {year} {2002})}\BibitemShut {NoStop}%
\bibitem [{\citenamefont {Jalaal}\ and\ \citenamefont
  {Balmforth}(2016)}]{jalaal2016long}%
  \BibitemOpen
  \bibfield  {author} {\bibinfo {author} {\bibfnamefont {M.}~\bibnamefont
  {Jalaal}}\ and\ \bibinfo {author} {\bibfnamefont {N.}~\bibnamefont
  {Balmforth}},\ }\href@noop {} {\bibfield  {journal} {\bibinfo  {journal}
  {Journal of Non-Newtonian Fluid Mechanics}\ }\textbf {\bibinfo {volume}
  {238}},\ \bibinfo {pages} {100} (\bibinfo {year} {2016})}\BibitemShut
  {NoStop}%
\bibitem [{\citenamefont {Labories}\ \emph {et~al.}(2017)\citenamefont
  {Labories}, \citenamefont {Rouyer}, \citenamefont {Angelescu},\ and\
  \citenamefont {Lorenceau}}]{jfm_laborie_2017}%
  \BibitemOpen
  \bibfield  {author} {\bibinfo {author} {\bibfnamefont {B.}~\bibnamefont
  {Labories}}, \bibinfo {author} {\bibfnamefont {F.}~\bibnamefont {Rouyer}},
  \bibinfo {author} {\bibfnamefont {D.}~\bibnamefont {Angelescu}},\ and\
  \bibinfo {author} {\bibfnamefont {E.}~\bibnamefont {Lorenceau}},\ }\href@noop
  {} {\bibfield  {journal} {\bibinfo  {journal} {J. Fluid Mech.}\ }\textbf
  {\bibinfo {volume} {818}},\ \bibinfo {pages} {838} (\bibinfo {year}
  {2017})}\BibitemShut {NoStop}%
\bibitem [{\citenamefont {Anna}\ \emph {et~al.}(2003)\citenamefont {Anna},
  \citenamefont {Bontoux},\ and\ \citenamefont {Stone}}]{apl_anna_2003}%
  \BibitemOpen
  \bibfield  {author} {\bibinfo {author} {\bibfnamefont {S.}~\bibnamefont
  {Anna}}, \bibinfo {author} {\bibfnamefont {N.}~\bibnamefont {Bontoux}},\ and\
  \bibinfo {author} {\bibfnamefont {H.}~\bibnamefont {Stone}},\ }\href@noop {}
  {\bibfield  {journal} {\bibinfo  {journal} {Appl. Phys. Lett.}\ }\textbf
  {\bibinfo {volume} {82}},\ \bibinfo {pages} {364} (\bibinfo {year}
  {2003})}\BibitemShut {NoStop}%
\bibitem [{\citenamefont {Garstecki}\ \emph {et~al.}(2006)\citenamefont
  {Garstecki}, \citenamefont {Fuerstman}, \citenamefont {Stone},\ and\
  \citenamefont {Whitesides}}]{m_garstecki_2006}%
  \BibitemOpen
  \bibfield  {author} {\bibinfo {author} {\bibfnamefont {P.}~\bibnamefont
  {Garstecki}}, \bibinfo {author} {\bibfnamefont {M.}~\bibnamefont
  {Fuerstman}}, \bibinfo {author} {\bibfnamefont {H.}~\bibnamefont {Stone}},\
  and\ \bibinfo {author} {\bibfnamefont {G.}~\bibnamefont {Whitesides}},\
  }\href@noop {} {\bibfield  {journal} {\bibinfo  {journal} {Lab Chip}\
  }\textbf {\bibinfo {volume} {6}},\ \bibinfo {pages} {437} (\bibinfo {year}
  {2006})}\BibitemShut {NoStop}%
\bibitem [{\citenamefont {Nisiako}\ \emph {et~al.}(2006)\citenamefont
  {Nisiako}, \citenamefont {Torii},\ and\ \citenamefont
  {Takizawa}}]{am_nisiako_2006}%
  \BibitemOpen
  \bibfield  {author} {\bibinfo {author} {\bibfnamefont {T.}~\bibnamefont
  {Nisiako}}, \bibinfo {author} {\bibfnamefont {T.}~\bibnamefont {Torii}},\
  and\ \bibinfo {author} {\bibfnamefont {Y.}~\bibnamefont {Takizawa}},\
  }\href@noop {} {\bibfield  {journal} {\bibinfo  {journal} {Adv. Mater.}\
  }\textbf {\bibinfo {volume} {18}},\ \bibinfo {pages} {1152} (\bibinfo {year}
  {2006})}\BibitemShut {NoStop}%
\bibitem [{\citenamefont {Tirandazi}\ and\ \citenamefont
  {Hidrovo}(2017)}]{tirandazi2017liquid}%
  \BibitemOpen
  \bibfield  {author} {\bibinfo {author} {\bibfnamefont {P.}~\bibnamefont
  {Tirandazi}}\ and\ \bibinfo {author} {\bibfnamefont {C.~H.}\ \bibnamefont
  {Hidrovo}},\ }\href@noop {} {\bibfield  {journal} {\bibinfo  {journal}
  {Journal of Micromechanics and Microengineering}\ }\textbf {\bibinfo {volume}
  {27}},\ \bibinfo {pages} {075020} (\bibinfo {year} {2017})}\BibitemShut
  {NoStop}%
\bibitem [{\citenamefont {Hirt}\ and\ \citenamefont
  {Nichols}(1981)}]{hirt1981volume}%
  \BibitemOpen
  \bibfield  {author} {\bibinfo {author} {\bibfnamefont {C.~W.}\ \bibnamefont
  {Hirt}}\ and\ \bibinfo {author} {\bibfnamefont {B.~D.}\ \bibnamefont
  {Nichols}},\ }\href@noop {} {\bibfield  {journal} {\bibinfo  {journal}
  {Journal of computational physics}\ }\textbf {\bibinfo {volume} {39}},\
  \bibinfo {pages} {201} (\bibinfo {year} {1981})}\BibitemShut {NoStop}%
\bibitem [{\citenamefont {Brackbill}\ \emph {et~al.}(1992)\citenamefont
  {Brackbill}, \citenamefont {Kothe},\ and\ \citenamefont
  {Zemach}}]{brackbill1992continuum}%
  \BibitemOpen
  \bibfield  {author} {\bibinfo {author} {\bibfnamefont {J.}~\bibnamefont
  {Brackbill}}, \bibinfo {author} {\bibfnamefont {D.~B.}\ \bibnamefont
  {Kothe}},\ and\ \bibinfo {author} {\bibfnamefont {C.}~\bibnamefont
  {Zemach}},\ }\href@noop {} {\bibfield  {journal} {\bibinfo  {journal}
  {Journal of computational physics}\ }\textbf {\bibinfo {volume} {100}},\
  \bibinfo {pages} {335} (\bibinfo {year} {1992})}\BibitemShut {NoStop}%
\bibitem [{\citenamefont {Abadie}\ \emph {et~al.}(2015)\citenamefont {Abadie},
  \citenamefont {Aubin},\ and\ \citenamefont {Legendre}}]{abadie2015combined}%
  \BibitemOpen
  \bibfield  {author} {\bibinfo {author} {\bibfnamefont {T.}~\bibnamefont
  {Abadie}}, \bibinfo {author} {\bibfnamefont {J.}~\bibnamefont {Aubin}},\ and\
  \bibinfo {author} {\bibfnamefont {D.}~\bibnamefont {Legendre}},\ }\href@noop
  {} {\bibfield  {journal} {\bibinfo  {journal} {Journal of Computational
  Physics}\ }\textbf {\bibinfo {volume} {297}},\ \bibinfo {pages} {611}
  (\bibinfo {year} {2015})}\BibitemShut {NoStop}%
\bibitem [{\citenamefont {Francois}\ \emph {et~al.}(2006)\citenamefont
  {Francois}, \citenamefont {Cummins}, \citenamefont {Dendy}, \citenamefont
  {Kothe}, \citenamefont {Sicilian},\ and\ \citenamefont
  {Williams}}]{francois2006balanced}%
  \BibitemOpen
  \bibfield  {author} {\bibinfo {author} {\bibfnamefont {M.~M.}\ \bibnamefont
  {Francois}}, \bibinfo {author} {\bibfnamefont {S.~J.}\ \bibnamefont
  {Cummins}}, \bibinfo {author} {\bibfnamefont {E.~D.}\ \bibnamefont {Dendy}},
  \bibinfo {author} {\bibfnamefont {D.~B.}\ \bibnamefont {Kothe}}, \bibinfo
  {author} {\bibfnamefont {J.~M.}\ \bibnamefont {Sicilian}},\ and\ \bibinfo
  {author} {\bibfnamefont {M.~W.}\ \bibnamefont {Williams}},\ }\href@noop {}
  {\bibfield  {journal} {\bibinfo  {journal} {Journal of Computational
  Physics}\ }\textbf {\bibinfo {volume} {213}},\ \bibinfo {pages} {141}
  (\bibinfo {year} {2006})}\BibitemShut {NoStop}%
\bibitem [{\citenamefont {Harvie}\ \emph {et~al.}(2006)\citenamefont {Harvie},
  \citenamefont {Davidson},\ and\ \citenamefont {Rudman}}]{harvie2006analysis}%
  \BibitemOpen
  \bibfield  {author} {\bibinfo {author} {\bibfnamefont {D.~J.}\ \bibnamefont
  {Harvie}}, \bibinfo {author} {\bibfnamefont {M.}~\bibnamefont {Davidson}},\
  and\ \bibinfo {author} {\bibfnamefont {M.}~\bibnamefont {Rudman}},\
  }\href@noop {} {\bibfield  {journal} {\bibinfo  {journal} {Applied
  mathematical modelling}\ }\textbf {\bibinfo {volume} {30}},\ \bibinfo {pages}
  {1056} (\bibinfo {year} {2006})}\BibitemShut {NoStop}%
\bibitem [{\citenamefont {Deshpande}\ \emph {et~al.}(2012)\citenamefont
  {Deshpande}, \citenamefont {Anumolu},\ and\ \citenamefont
  {Trujillo}}]{deshpande2012evaluating}%
  \BibitemOpen
  \bibfield  {author} {\bibinfo {author} {\bibfnamefont {S.~S.}\ \bibnamefont
  {Deshpande}}, \bibinfo {author} {\bibfnamefont {L.}~\bibnamefont {Anumolu}},\
  and\ \bibinfo {author} {\bibfnamefont {M.~F.}\ \bibnamefont {Trujillo}},\
  }\href@noop {} {\bibfield  {journal} {\bibinfo  {journal} {Computational
  science \& discovery}\ }\textbf {\bibinfo {volume} {5}},\ \bibinfo {pages}
  {014016} (\bibinfo {year} {2012})}\BibitemShut {NoStop}%
\bibitem [{\citenamefont {Sign\'{e}~Mamba}\ \emph {et~al.}(2019)\citenamefont
  {Sign\'{e}~Mamba}, \citenamefont {Magniez}, \citenamefont {Zoueshtiagh},\
  and\ \citenamefont {Baudoin}}]{ijmf_signe_2019}%
  \BibitemOpen
  \bibfield  {author} {\bibinfo {author} {\bibfnamefont {S.}~\bibnamefont
  {Sign\'{e}~Mamba}}, \bibinfo {author} {\bibfnamefont {J.}~\bibnamefont
  {Magniez}}, \bibinfo {author} {\bibfnamefont {F.}~\bibnamefont
  {Zoueshtiagh}},\ and\ \bibinfo {author} {\bibfnamefont {M.}~\bibnamefont
  {Baudoin}},\ }\href@noop {} {\bibfield  {journal} {\bibinfo  {journal} {Int.
  J. Multiph. Flow}\ }\textbf {\bibinfo {volume} {113}},\ \bibinfo {pages}
  {343} (\bibinfo {year} {2019})}\BibitemShut {NoStop}%
\bibitem [{\citenamefont {Fujioka}\ and\ \citenamefont
  {Grotberg}(2005)}]{fujioka2005steady}%
  \BibitemOpen
  \bibfield  {author} {\bibinfo {author} {\bibfnamefont {H.}~\bibnamefont
  {Fujioka}}\ and\ \bibinfo {author} {\bibfnamefont {J.~B.}\ \bibnamefont
  {Grotberg}},\ }\href@noop {} {\bibfield  {journal} {\bibinfo  {journal}
  {Physics of Fluids}\ }\textbf {\bibinfo {volume} {17}},\ \bibinfo {pages}
  {082102} (\bibinfo {year} {2005})}\BibitemShut {NoStop}%
\bibitem [{\citenamefont {Fujioak}\ \emph {et~al.}(2008)\citenamefont
  {Fujioak}, \citenamefont {Takayama},\ and\ \citenamefont
  {Grotberg}}]{pof_fujioka_2008}%
  \BibitemOpen
  \bibfield  {author} {\bibinfo {author} {\bibfnamefont {H.}~\bibnamefont
  {Fujioak}}, \bibinfo {author} {\bibfnamefont {S.}~\bibnamefont {Takayama}},\
  and\ \bibinfo {author} {\bibfnamefont {J.}~\bibnamefont {Grotberg}},\
  }\href@noop {} {\bibfield  {journal} {\bibinfo  {journal} {Phys. Fluids}\
  }\textbf {\bibinfo {volume} {20}},\ \bibinfo {pages} {062104} (\bibinfo
  {year} {2008})}\BibitemShut {NoStop}%
\bibitem [{\citenamefont {Hassan}\ \emph {et~al.}(2011)\citenamefont {Hassan},
  \citenamefont {Uzgoren}, \citenamefont {Fujioka}, \citenamefont {Grotberg},\
  and\ \citenamefont {Shyy}}]{hassan2011adaptive}%
  \BibitemOpen
  \bibfield  {author} {\bibinfo {author} {\bibfnamefont {E.~A.}\ \bibnamefont
  {Hassan}}, \bibinfo {author} {\bibfnamefont {E.}~\bibnamefont {Uzgoren}},
  \bibinfo {author} {\bibfnamefont {H.}~\bibnamefont {Fujioka}}, \bibinfo
  {author} {\bibfnamefont {J.~B.}\ \bibnamefont {Grotberg}},\ and\ \bibinfo
  {author} {\bibfnamefont {W.}~\bibnamefont {Shyy}},\ }\href@noop {} {\bibfield
   {journal} {\bibinfo  {journal} {International journal for numerical methods
  in fluids}\ }\textbf {\bibinfo {volume} {67}},\ \bibinfo {pages} {1373}
  (\bibinfo {year} {2011})}\BibitemShut {NoStop}%
\bibitem [{\citenamefont {Jensen}(2000)}]{jcis_jensen_2000}%
  \BibitemOpen
  \bibfield  {author} {\bibinfo {author} {\bibfnamefont {O.}~\bibnamefont
  {Jensen}},\ }\href@noop {} {\bibfield  {journal} {\bibinfo  {journal} {J.
  Coll. Int. Sci.}\ }\textbf {\bibinfo {volume} {221}},\ \bibinfo {pages} {38}
  (\bibinfo {year} {2000})}\BibitemShut {NoStop}%
\bibitem [{\citenamefont {Ratulowski}\ and\ \citenamefont
  {Chang}(1989)}]{ratulowski1989transport}%
  \BibitemOpen
  \bibfield  {author} {\bibinfo {author} {\bibfnamefont {J.}~\bibnamefont
  {Ratulowski}}\ and\ \bibinfo {author} {\bibfnamefont {H.-C.}\ \bibnamefont
  {Chang}},\ }\href@noop {} {\bibfield  {journal} {\bibinfo  {journal} {Physics
  of Fluids A: Fluid Dynamics}\ }\textbf {\bibinfo {volume} {1}},\ \bibinfo
  {pages} {1642} (\bibinfo {year} {1989})}\BibitemShut {NoStop}%
\bibitem [{\citenamefont {Ajaev}\ and\ \citenamefont
  {Homsy}(2006)}]{ajaev2006modeling}%
  \BibitemOpen
  \bibfield  {author} {\bibinfo {author} {\bibfnamefont {V.~S.}\ \bibnamefont
  {Ajaev}}\ and\ \bibinfo {author} {\bibfnamefont {G.}~\bibnamefont {Homsy}},\
  }\href@noop {} {\bibfield  {journal} {\bibinfo  {journal} {Annu. Rev. Fluid
  Mech.}\ }\textbf {\bibinfo {volume} {38}},\ \bibinfo {pages} {277} (\bibinfo
  {year} {2006})}\BibitemShut {NoStop}%
\bibitem [{\citenamefont {Hoffman}(1975)}]{hoffman1975study}%
  \BibitemOpen
  \bibfield  {author} {\bibinfo {author} {\bibfnamefont {R.~L.}\ \bibnamefont
  {Hoffman}},\ }\href@noop {} {\bibfield  {journal} {\bibinfo  {journal}
  {Journal of colloid and interface science}\ }\textbf {\bibinfo {volume}
  {50}},\ \bibinfo {pages} {228} (\bibinfo {year} {1975})}\BibitemShut
  {NoStop}%
\bibitem [{\citenamefont {Boender}\ \emph {et~al.}(1991)\citenamefont
  {Boender}, \citenamefont {Chesters},\ and\ \citenamefont {Van
  Der~Zanden}}]{boender1991approximate}%
  \BibitemOpen
  \bibfield  {author} {\bibinfo {author} {\bibfnamefont {W.}~\bibnamefont
  {Boender}}, \bibinfo {author} {\bibfnamefont {A.}~\bibnamefont {Chesters}},\
  and\ \bibinfo {author} {\bibfnamefont {A.}~\bibnamefont {Van Der~Zanden}},\
  }\href@noop {} {\bibfield  {journal} {\bibinfo  {journal} {International
  journal of multiphase flow}\ }\textbf {\bibinfo {volume} {17}},\ \bibinfo
  {pages} {661} (\bibinfo {year} {1991})}\BibitemShut {NoStop}%
\bibitem [{\citenamefont {Landau}\ and\ \citenamefont
  {Levich}(1942)}]{landau2012dragging}%
  \BibitemOpen
  \bibfield  {author} {\bibinfo {author} {\bibfnamefont {L.}~\bibnamefont
  {Landau}}\ and\ \bibinfo {author} {\bibfnamefont {B.}~\bibnamefont
  {Levich}},\ }\href@noop {} {\bibfield  {journal} {\bibinfo  {journal} {Acta
  Physicochimica URSS}\ ,\ \bibinfo {pages} {42}} (\bibinfo {year}
  {1942})}\BibitemShut {NoStop}%
\end{thebibliography}

%

\end{document}